\documentclass{appolb}
\usepackage{graphicx}
\usepackage{amsmath}
\usepackage{mathrsfs}
\usepackage[pdftex,colorlinks,bookmarksnumbered]{hyperref}   
\hypersetup{citecolor=[rgb]{0,0.4,0.8},linkcolor=red,urlcolor=[rgb]{0.5,0,0}}

\begin{document}
\title{TOROIDAL MODES IN NUCLEI BY INELASTIC ELECTRON SCATTERING
\thanks{Joint contribution from the talks of Repko and Kvasil presented at the XXV Nuclear Physics Workshop ``Marie and Pierre Curie'', Kazimierz Dolny, Poland, September 25--30, 2018.}%
}
\author{Anton Repko
\address{Institute of Physics, Slovak Academy of Sciences, 84511 Bratislava, Slovakia}
\\[10pt]
{Jan Kvasil
}
\address{IPNP, Math.-Phys. Faculty, Charles University, 18000 Praha 8, Czech Republic}
}
\maketitle
\begin{abstract}
Electron scattering is a tool that can provide relatively clean view of the nuclear structure in both ground and excited states, as it depends on the well-known electromagnetic interaction. But since the common expressions for its cross section were derived with certain assumptions, in this paper we describe several nontrivial steps necessary for a proper theoretical calculation within the current density-functional framework, namely with Skyrme QRPA for axial nuclei, with aim to enable comparison of the theoretically predicted low-lying $1^-$ toroidal modes with future $(e,e')$ experiments.
\end{abstract}

\section{Introduction}
\begin{figure}[htb]
\centerline{%
\includegraphics[width=10cm]{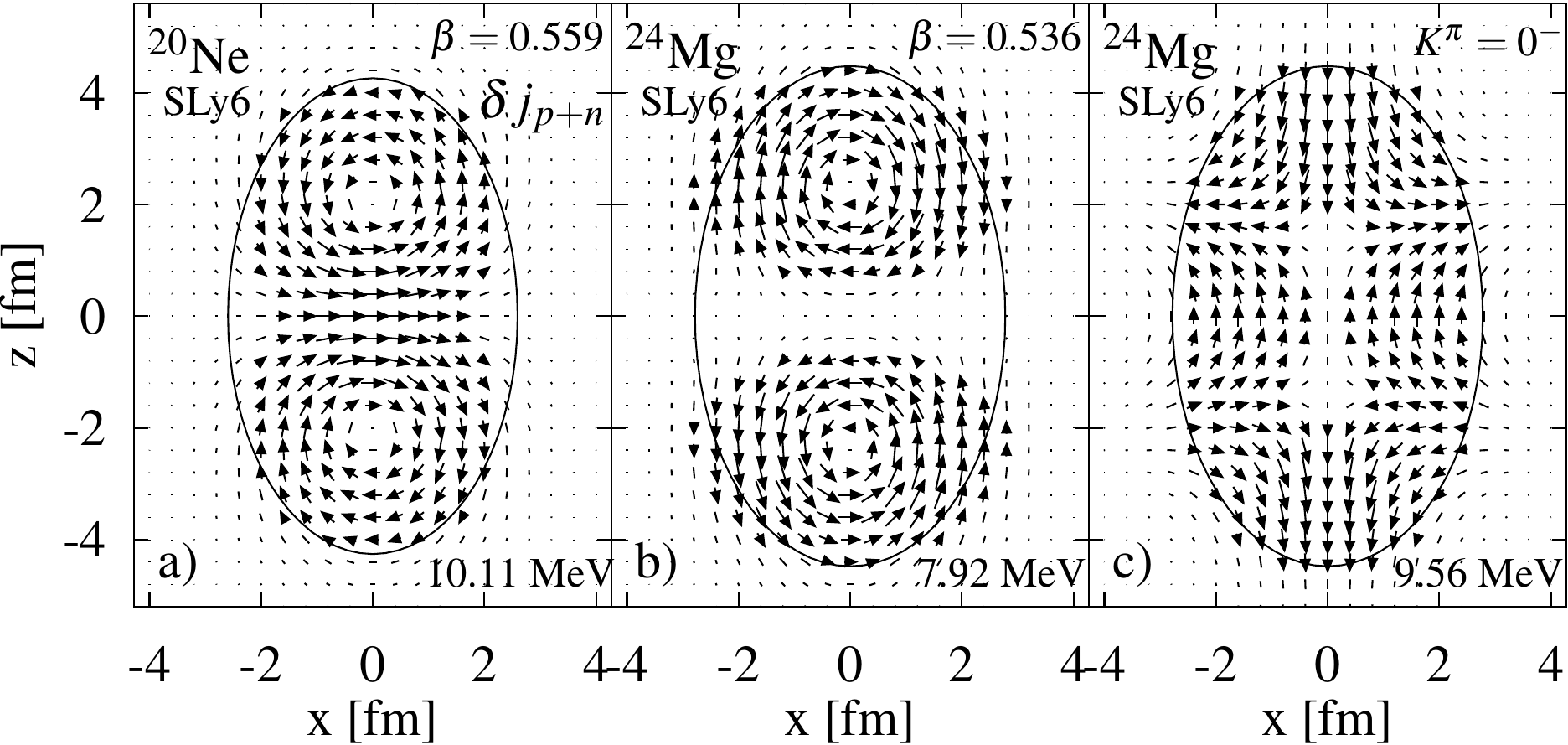}}
\caption{Plots of the isoscalar transition convective current in the low-lying $K^{\pi}=1^-$ states in a) $^{20}$Ne, b) $^{24}$Mg with toroid-like (vortical) character, and c) $K^{\pi}=0^-$ state in $^{24}$Mg which has a compression admixture.}
\label{Fig:tor_cur}
\end{figure}

Folowing our recent study of $^{208}$Pb \cite{Re13}, the usual interpretation of pygmy $E1$ mode \cite{Sa13} can be questioned in favor of isoscalar toroidal mode. Besides, there are predictions that low-lying $1^-$ states in light nuclei $^{10}$Be \cite{Ka17}, $^{24}$Mg \cite{Ne18}, $^{20}$Ne \cite{Burgas} have mainly vortical character (Fig.~\ref{Fig:tor_cur}). Inelastic electron scattering, which is a long-known method in the study of nuclear structure \cite{FW66,HB83}, appears as an ideal tool to resolve these questions on the state-by-state basis, mainly in the light nuclei, characterized by low density of excitations. In contrast, the strong-interaction-dependent $\alpha$-scattering probes mainly the transition densities, related to the compression current; whereas the toroidal current is decoupled from the density, and therefore supressed in $(\alpha,\alpha')$.

In order to provide a reliable theoretical treatment of the $(e,e')$ scattering in connection with Skyrme functional, we discuss: (i) removal of the spurious modes, (ii) effective current for Skyrme, (iii) exact relativistic kinematics, (iv) adaptation of the form-factor formalism from spherical to axial nuclei, and (v) other known corrections, namely the recoil term and the effective momentum, which mimics the distorted-wave Born approximation (DWBA), while we are still working with the less-demanding plane-wave Born approximation (PWBA). The above-mentioned topics address namely: (i) correct structure of the low-energy states, (ii) mutual proportion of longitudinal and transversal form-factors, and (iii) the area of low transferred momentum, crucial for the discernment of exotic modes. Although in principle we should be able to reconstruct the transition densities and currents by the Fourier-Bessel transformation of the measured data \cite{HB83}, the corrections mentioned above are still needed for a proper analysis.

\section{Overview of the Skyrme QRPA and spurious removal}
We are working in the framework of axial quasiparticle random phase approximation (QRPA) on top of Skyrme density functional \cite{Re15}. Phonons are constructed with a well defined angular momentum $\mu$ (also denoted as $K=|\mu|$) and parity $\pi$, and are numbered by index $\nu$:
\begin{equation}
\hat{Q}_\nu^+ = \!\!\!\sum_{i>j}^{K_i+K_j=\mu}\!\!\! \Big(\mathcal{X}_{ij}^{(\nu)} \hat{\alpha}_i^+\hat{\alpha}_{j}^+ - \mathcal{Y}_{ij}^{(\nu)}\hat{\alpha}_{\bar{j}}\hat{\alpha}_{\bar{i}}\Big)
\end{equation}
Solution of the QRPA equation $[\hat{H},\hat{Q}_\nu^+] = \hbar\omega_\nu\hat{Q}_\nu^+$ is then formulated as a matrix eigenproblem
\begin{equation}
\begin{pmatrix} \mathbf{A} & \mathbf{B} \\ \mathbf{B} & \mathbf{A} \end{pmatrix}
\begin{pmatrix} \mathcal{X}^{(\nu)} \\ \mathcal{Y}^{(\nu)} \end{pmatrix} =
\hbar\omega_\nu \begin{pmatrix} \mathcal{X}^{(\nu)} \\ -\mathcal{Y}^{(\nu)} \end{pmatrix}
\end{equation}
where matrices $\mathbf{A},\ \mathbf{B}$ are composed of the second functional derivatives of the Skyrme functional. Transition density and current are then calculated as ground-state commutators, by assuming $Q_\nu|0\rangle = 0$.
\begin{equation}
\delta\rho_\nu(\vec{r}_{\,}) \equiv \langle\nu|\hat{\rho}(\vec{r})|0\rangle = \langle[\hat{Q}_\nu,\hat{\rho}(\vec{r})]\rangle,
\quad \delta\vec{j}_\nu(\vec{r}) \equiv \langle\nu|\hat{\vec{j}}(\vec{r})|0\rangle = \langle[\hat{Q}_\nu,\hat{\vec{j}}(\vec{r})]\rangle
\end{equation}

In the case of spontaneously broken symmetry (translation, rotation, particle conservation), QRPA spectrum will contain unphysical spurious states of certain multipolarity: center-of-mass motion appears as $I^\pi = 1^-$ (axial nuclei: $K^\pi=0^-$ or $1^-$), rotation of an axially deformed nucleus as $K^\pi = 1^+$, and particle-number violation by pairing (separately for protons and neutrons) as $K^\pi=0^+$ states. In an ideal case (complete basis), these spurious states appear in pairs as generalized eigenstates, defined by a time-even operator $\hat{X}$ and a time-odd symmetry generator $\hat{P}$ \cite{Re18},
\begin{equation}
\hat{X} = \!\!\!\sum_{i>j}^{K_i+K_j=\mu}\!\!\! {X}_{ij} (\hat{\alpha}_i^+\hat{\alpha}_{j}^+ + \hat{\alpha}_{\bar{j}}\hat{\alpha}_{\bar{i}}),\quad
\hat{P} = \!\!\!\sum_{i>j}^{K_i+K_j=\mu}\!\!\! {P}_{ij} (\hat{\alpha}_i^+\hat{\alpha}_{j}^+ - \hat{\alpha}_{\bar{j}}\hat{\alpha}_{\bar{i}}),
\end{equation}
which are expected to fulfill the corresponding QRPA equations
\begin{subequations}
\begin{align}
{[\hat{H},\hat{P}]} &= 0 &\Rightarrow\quad
&\begin{pmatrix} \mathbf{A} & \mathbf{B} \\ \mathbf{B} & \mathbf{A} \end{pmatrix}
\begin{pmatrix} P \\ P \end{pmatrix} = 0 \\
{[\hat{H},\hat{X}]} &= -\mathrm{i}\hat{P} &\Rightarrow\quad
&\begin{pmatrix} \mathbf{A} & \mathbf{B} \\ \mathbf{B} & \mathbf{A} \end{pmatrix}
\begin{pmatrix} X \\ -X \end{pmatrix} = -\mathrm{i}
\begin{pmatrix} P \\ -P \end{pmatrix} \\
&&\textrm{so}\quad&X_{ij} = -\mathrm{i}{\big[(\mathbf{A}-\mathbf{B})^{-1}P\big]}_{ij}.
\end{align}
\end{subequations}
Since the real calculation with finite basis doesn't separate the spurious modes exactly, their admixture needs to be projected out from solutions $\nu$, to get corrected ones ($\nu'$), by requiring $\langle[Q_{\nu'},\hat{X}]\rangle = \langle[Q_{\nu'},\hat{P}]\rangle = 0$.
\begin{equation}
\hat{Q}_{\nu'}^+ = \hat{Q}_\nu^+
- \frac{\langle[\hat{P}^\dagger,\hat{Q}_\nu^+]\rangle}{\langle[\hat{P}^\dagger,\hat{X}]\rangle}\hat{X}
- \frac{\langle[\hat{X}^\dagger,\hat{Q}_\nu^+]\rangle}{\langle[\hat{X}^\dagger,\hat{P}]\rangle}\hat{P}
\end{equation}
For the spurious pairing state ($0^+$), the role of $\hat{X}$ and $\hat{P}$ is swapped. Further details can be found elsewhere \cite{Re18,Na17}.

\subsection{Effective current}
It turns out that the continuity equation is not fulfilled for the convective current operator, defined here according to Skyrme DFT notation ($q=p,n$)
\begin{equation}
\label{jbare}
\hat{\vec{j}}_q(\vec{r}) = \frac{\mathrm{i}}{2}\sum_{j\in q} \Big[\overleftarrow{\nabla}_j\delta^3(\vec{r}_j-\vec{r}_{\,})
 - \delta^3(\vec{r}_j-\vec{r}_{\,})\overrightarrow{\nabla}_j\Big],
\end{equation}
and therefore gives slightly wrong scaling for the quantities, which depend on isovector transition current, namely the toroidal transition probabilities and transversal form-factors for $e^-$ scattering. The problem can be briefly clarified by saying that the bare current (\ref{jbare}) deals with the momentum density, whereas we need ``velocity'' (or electric current) density; so the effective mass should be somehow involved. The magnetization current is not altered.

We will outline here a heuristic derivation of $\hat{\vec{j}}_{\mathrm{eff}}(\vec{r})$, the proper effective current,  by evaluating the time evolution of the center-of-mass coordinate in Heisenberg picture as $\partial\hat{X}/\partial t = \mathrm{i}/\hbar [\hat{H},\hat{X}]$. By working in tensor-operator formalism and employing the units of (\ref{jbare}), we assert (for each component $\mu$)
\begin{equation}
\hat{X}_{q;\mu} = \vec{e}_\mu\cdot\sum_{i\in q}\vec{r}_i = \sqrt{\tfrac{4\pi}{3}}\sum_{i\in q}r_i Y_{1\mu}(\hat{r}_i), \ \ 
[\hat{H},\hat{X}_{q;\mu}] = \tfrac{-\mathrm{i}\hbar^2}{m_q}\int \vec{e}_\mu\cdot\hat{\vec{j}}_{\mathrm{eff};q}(\vec{r})\,\mathrm{d}^3 r.
\end{equation}
After inserting the Skyrme Hamiltonian (containing first and second functional derivatives) and ``localization'' of the integral, we finally obtain
\begin{align}
\hat{\vec{j}}_{\mathrm{eff};q}(\vec{r}) &= \hat{\vec{j}}_q(\vec{r})
+\frac{m_q}{\hbar^2}\Big\{
2b_1\big[\rho_{\bar{q}}(\vec{r})\hat{\vec{j}}_q(\vec{r})
- \rho_q(\vec{r})\hat{\vec{j}}_{\bar{q}}(\vec{r})\big] \nonumber\\
\label{jeff}
&\quad{}+b_4\big[\rho_{\bar{q}}(\vec{r})\vec{\nabla}\times\hat{\vec{\sigma}}_q(\vec{r})
-\rho_q(\vec{r})\vec{\nabla}\times\hat{\vec{\sigma}}_{\bar{q}}(\vec{r})\big]
+\mathcal{J}\textrm{-terms}\Big\},
\end{align}
and this current fulfills the continuity equation in the given units
\begin{equation}
\label{cont_eq}
\vec{\nabla}\cdot\hat{\vec{j}}_{\mathrm{eff};q}(\vec{r})
= -\frac{m_q}{\hbar}\frac{\partial\hat{\rho}_q(\vec{r})}{\partial t},
\quad\textrm{where}\quad
\hat{\rho}_q(\vec{r}) = \sum_{j\in q} \delta^3(\vec{r}_j-\vec{r}_{\,}).
\end{equation}
We used $\bar{q}$ to denote the opposite particle type ($p\leftrightarrow n$), and $\mathcal{J}$-terms represent a complicated spin-orbital contribution of the usually neglected $\mathcal{J}^2$ part of the Skyrme functional. These results bear some similarity to the derivation of the isovector EWSR (energy-weighted sum rule) in \cite{Li89}. It is also clear that the corrections in (\ref{jeff}) disappear for the isoscalar excitations.

\begin{figure}[htb]
\centerline{%
\includegraphics[width=12.5cm]{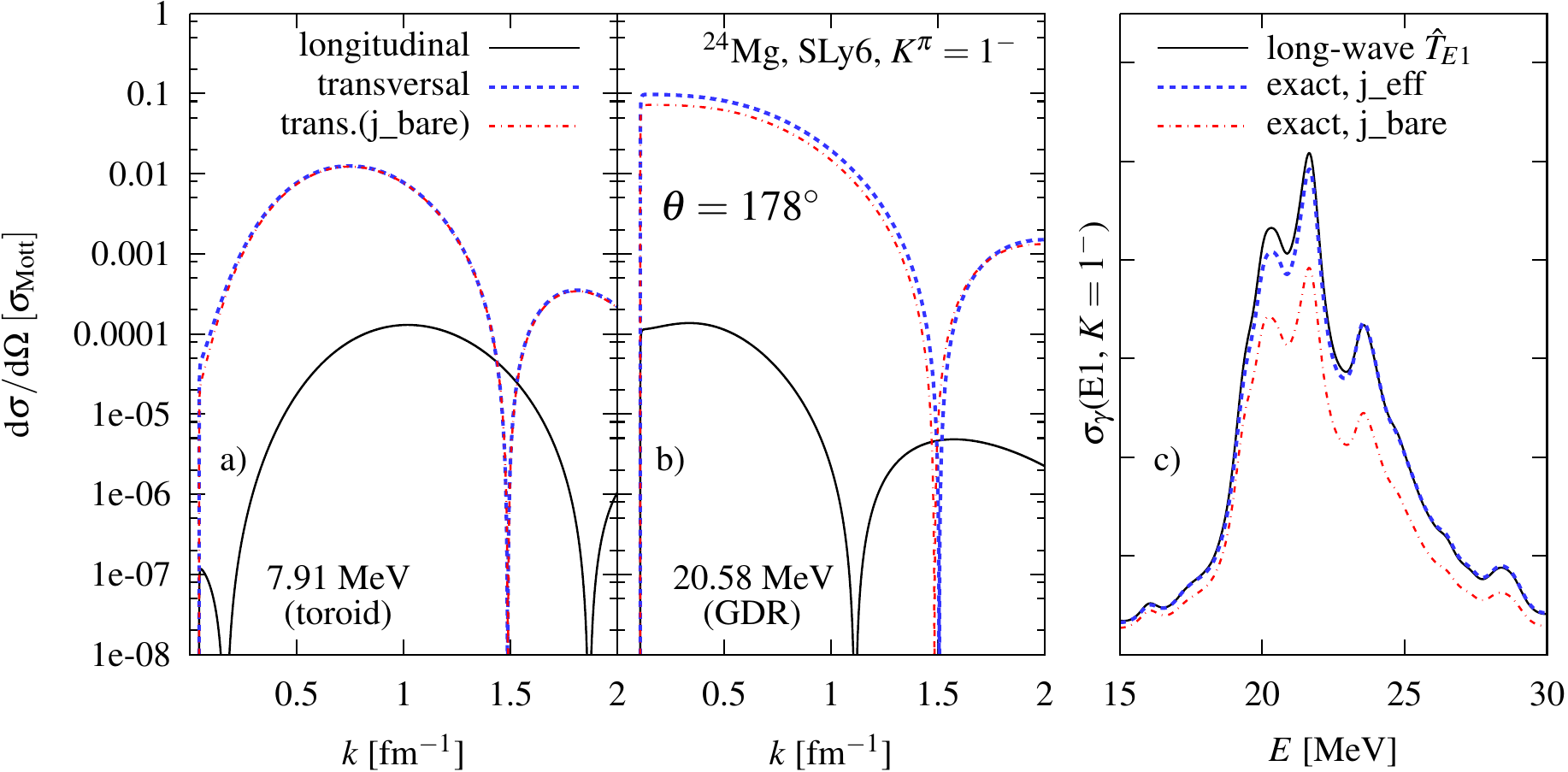}}
\caption{Plots showing the difference of bare and effective current. Left pair of plots: longitudinal (density-dependent) and transversal (current-dependent) part of the $\mathrm{d}\sigma/\mathrm{d}\Omega$ for electron scattering at a) toroidal isoscalar state, b) GDR-like isovector state. Panel c) shows the E1 photoabsorption strength function (component $K^\pi=1^-$ with smoothing $\Delta=1\ \mathrm{MeV}$) calculated with long-wave (density-dependent) and exact (current-dependent) transition operator.}
\label{Fig:eff_cur}
\end{figure}

To illustrate the influence of the effective current, we are showing here (Fig.~\ref{Fig:eff_cur}) the plots of longitudinal and transversal electron-scattering cross sections (explained below) and also photoabsorption strength function. The ``exact'' strength function involves the full electric transition operator of given multipolarity, and is compared to usual long-wave approximation. 
\begin{subequations}
\begin{align}
\hat{T}_{E\lambda\mu}^\mathrm{longw} &=
\int e_p\hat{\rho}_p(\vec{r})
r^\lambda Y_{\lambda\mu}(\vartheta,\varphi)\,\mathrm{d}^3 r \\
\hat{T}_{E\lambda\mu}^\mathrm{exact} &=
-\frac{(2\lambda+1)!!}{ck^{\lambda+1}} \sqrt{\frac{\lambda}{\lambda+1}}
\int \hat{\vec{j}}_\mathrm{nuc}(\vec{r})\cdot\vec{\nabla}\times
\big[j_\lambda(kr)\vec{Y}_{\lambda\mu}^\lambda(\vartheta,\varphi)\big]\,\mathrm{d}^3 r \\
\label{j_nuc}
&\!\!\!\textrm{where }\,k=\tfrac{\omega}{c},\ \, 
\hat{\vec{j}}_\mathrm{nuc}(\vec{r}) = \sum_q \Big\{\frac{e_q\hbar}{m_p}
\hat{\vec{j}}_{\mathrm{eff};q}(\vec{r})
+ \mu_N\frac{g_{s;q}}{2}\vec{\nabla}\times\vec{\sigma}_q(\vec{r}) \Big\}
\end{align}
\end{subequations}

Fig.~\ref{Fig:eff_cur} also shows the main difference between collective vibration (GDR) and toroidal motion -- the decreasing form-factors at low $k$ for the latter -- which may be understood in terms of Fourier transform, where the low-momentum part reflects the spatial average (which cancels for curly motion).

\section{Inelastic electron scattering}
\begin{figure}[htb]
\centerline{%
\includegraphics{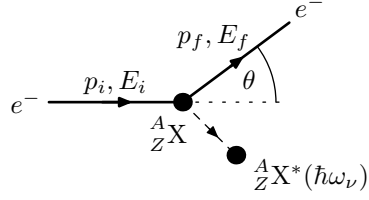}}
\caption{Depiction of the geometry of inelastic electron scattering in the laboratory frame. Static nucleus $^{A}_Z X$ gets excited to energy $\hbar\omega_\nu$, and receives momentum $q$.}
\label{Fig:el_kin}
\end{figure}
Inelastic electron scattering from nuclei is usually performed with stationary target nucleus (with mass $M$) and incident electrons accelerated to tens or hundreds MeV. Scattered electrons at given angle $\theta$ are detected, and their energy is measured (see Fig.~\ref{Fig:el_kin}), which enables to identify the nuclear excitation energy $\hbar\omega_\nu$.  Momentum transfer is denoted $\vec{q}=\vec{p}_i-\vec{p}_f$, and is often expressed in terms of wave-vector magnitude $k$ [fm$^{-1}$].
\begin{equation}
\hbar k = q = |\vec{q}_{\,}| = |\vec{p}_i-\vec{p}_f|
\end{equation}
Energy transfer is denoted $\Delta E$, and by employing $a-b=\tfrac{a^2-b^2}{a+b}$ (the formula used frequently also in Sec.~\ref{sec:rel}) we obtain
\begin{equation}
\label{DeltaE}
\Delta E \equiv E_i - E_f = E_n - Mc^2 = \hbar\omega_\nu + \frac{q^2}{E_n/c^2 + M + \hbar\omega_\nu/c^2}
\end{equation}
where we denote the total energy of the recoiled nucleus as
\begin{equation}
E_n = \sqrt{(Mc^2+\hbar\omega_\nu)^2 + q^2 c^2}.
\end{equation}
Another often employed quantity is the squared four-momentum transfer $Q^2$, which is always positive, as follows later from (\ref{x_def}).
\begin{equation}
\label{Q2}
Q^2 \equiv q^2-\tfrac{(\Delta E)^2}{c^2}
= 2\Big(\tfrac{E_i E_f}{c^2} - p_i p_f\cos\theta - m_e^2 c^2\Big)
\end{equation}

Differential cross section of the $e^-$ scattering (in PWBA approximation) can be expressed in terms of longitudinal (Coulomb; $F_{\lambda;fi}^c$) and transversal form-factor \cite{FW66,HB83}. Transversal form-factor is either electric ($F_{\lambda;fi}^e$) or magnetic ($F_{\lambda;fi}^m$), depending on the multipolarity and parity of the transition.
\begin{align}
\label{crosssec}
\frac{\mathrm{d}\sigma_{fi}}{\mathrm{d}\Omega}
&= \frac{8\pi\alpha^2\hbar^2}{e^2 c^2}f_\mathrm{rec}\frac{p_f}{p_i}
\bigg[\frac{E_i E_f + \vec{p}_i\cdot\vec{p}_f c^2+m_e^2 c^4}{q^4}
\big|F_{\lambda;fi}^c(k)\big|^2 \nonumber\\
&\qquad{}+
\bigg(\frac{p_i^2 p_f^2 c^2}{q^2 Q^4}\sin^2\theta + \frac{c^2}{2Q^2}\bigg)
\Big(\big|F_{\lambda;fi}^e(k)\big|^2 + \big|F_{\lambda;fi}^m(k)\big|^2\Big)\bigg] \\
\label{crosssec_approx}
&\approx \frac{\pi(\alpha\hbar c)^2}{e^2 E_i^2}
f_\mathrm{rec}\frac{\cos^2\tfrac{\theta}{2}}{\sin^4\tfrac{\theta}{2}}
\Big[\tfrac{Q^4}{q^4}\big|F_{\lambda;fi}^c(k)\big|^2 +
\Big(\tfrac{Q^2}{2q^2}\!+\!\tan^2\tfrac{\theta}{2}\Big)
\big|F_{\lambda;fi}^{e,m}(k)\big|^2\Big]
\end{align}
where $f_\mathrm{rec}$ is the recoil term (described in Sec.~\ref{sec:corr}). Eq.~(\ref{crosssec_approx}) gives a standard approximation with zero electron mass (analyzed in Sec.~\ref{sec:rel}). The form-factors for spherical nuclei are defined in terms of reduced matrix elements
\begin{subequations}\label{ff_sph}
\begin{align}
F_{\lambda;fi}^{c,e,m}(k) &= \frac{1}{\sqrt{2J_i+1}} \langle f|| \hat{T}_\lambda^{c,e,m}(k) ||i\rangle,
\quad\textrm{where}\\
&\hat{T}_{\lambda\mu}^c(k) = \sum_{q=p,n} e_q
 \int\hat{\rho}_q(\vec{r}) j_\lambda(kr)Y_{\lambda\mu}(\hat{r})\,\mathrm{d}^3 r \\
&\hat{T}_{\lambda\mu}^e(k) = \frac{\mathrm{i}}{ck}\int\hat{\vec{j}}_\mathrm{nuc}(\vec{r})\cdot
\big[\vec{\nabla}\times j_\lambda(kr)\vec{Y}_{\lambda\mu}^\lambda(\hat{r})\big]\mathrm{d}^3 r \\
&\hat{T}_{\lambda\mu}^m(k) = \frac{\mathrm{i}}{c}\int\hat{\vec{j}}_\mathrm{nuc}(\vec{r})\cdot
j_\lambda(kr)\vec{Y}_{\lambda\mu}^\lambda(\hat{r})\,\mathrm{d}^3 r, \\
&\quad\ \textrm{see }
\hat{\rho}_q(\vec{r})\ (\ref{cont_eq}),\,\hat{\vec{j}}_\mathrm{nuc}(\vec{r})\ (\ref{j_nuc});\quad
e_p=|e|,\,e_n=0,\,g_s=0.7g_{s,\mathrm{free}}. \nonumber
\end{align}
\end{subequations}

\subsection{\texorpdfstring{Exact relativistic kinematics for fixed $\theta$}{Exact relativistic kinematics for fixed theta}}\label{sec:rel}
When calculating scattering plots such as in Fig.~\ref{Fig:eff_cur}, we need to calculate $E_i, E_f$ in terms of given $\hbar\omega_\nu,\,q$ and $\theta$, which is rather nontrivial to do exactly. First, we directly obtain $\Delta E$ (\ref{DeltaE}) and $Q^2$ (\ref{Q2}). Then, we define quantity $x$, and use it to decompose $\tfrac{1}{2}Q^2$ (\ref{Q2}) by employing $\cos\theta=\cos^2\tfrac{\theta}{2}-\sin^2\tfrac{\theta}{2}$.
\begin{align}
\label{x_def}
x &\equiv \tfrac{E_i E_f}{c^2} - m_e^2 c^2 - p_i p_f
= \tfrac{m_e^2(E_i-E_f)^2 c^2}{E_i E_f + p_i p_f c^2 - m_e^2c^4}
= \tfrac{m_e^2(p_i-p_f)^2 c^4}{E_i E_f + p_i p_f c^2 + m_e^2c^4} \\
\tfrac{1}{2}Q^2 &= \Big(\tfrac{E_i E_f}{c^2} - p_i p_f - m_e^2 c^2\Big)\cos^2\tfrac{\theta}{2}
+ \Big(\tfrac{E_i E_f}{c^2} + p_i p_f - m_e^2 c^2\Big)\sin^2\tfrac{\theta}{2} \nonumber\\
&= x \cos^2\tfrac{\theta}{2} + \tfrac{(m_e\Delta E)^2}{x}\sin^2\tfrac{\theta}{2}
\end{align}
Thus we obtained a quadratic equation for $x$, with an interesting by-product of a second, low-$E_i$ (large-$x$) solution for $\theta<90^\circ$, which we can ignore due to a very low cross section associated with it. The usual solution reads
\begin{equation}
x = \frac{4(m_e\Delta E)^2\sin^2\tfrac{\theta}{2}}{Q^2+\sqrt{Q^4-4(m_e\Delta E)^2\sin^2\theta}},
\end{equation}
and, finally, for the best numerical accuracy, we can proceed by calculating $p_f$ from another quadratic equation, defined by $p_i p_f$ and $\Delta p = p_i-p_f$, which are evaluated by manipulating (\ref{x_def}) and (\ref{Q2}).
\begin{equation}
\Delta p = \sqrt{\frac{(\Delta E)^2}{c^2} + 2x},\ \quad
p_i p_f = \frac{Q^2\cos\theta + \sqrt{Q^4-4(m_e\Delta E)^2\sin^2\theta}}{2\sin^2\theta}
\end{equation}

\begin{figure}[htb]
\centerline{%
\includegraphics[width=12.5cm]{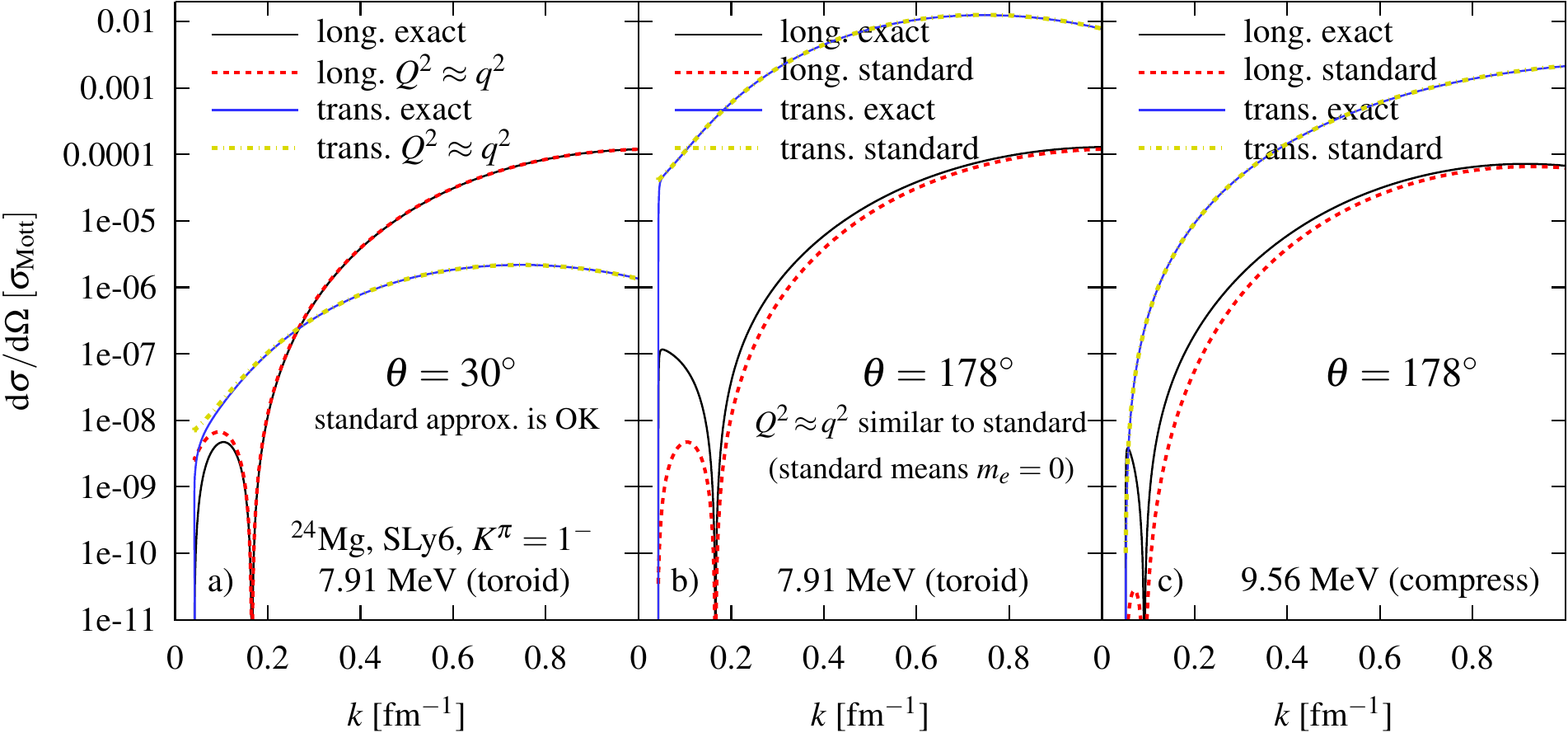}}
\caption{The impact of kinematic approximations for the forward and backward scattering on the differential cross section. Standard approximation ($m_e=0$) affects mainly the longitudinal back-scattering cross section at low $k$, while the additional $Q^2\approx q^2$ in forward scattering slightly affects also the transversal part.}
\label{Fig:relativ}
\end{figure}

The results obtained with exact relativistic kinematics and cross-section formula (\ref{crosssec}) are compared to standard approximation ($m_e=0$) and to another approximation, $Q^2 \approx q^2$, in Fig.~\ref{Fig:relativ}.

\subsection{Form-factor formulation for axial nuclei}\label{sec_axial}
Most nuclei are deformed, and their structure can be readily calculated by Skyrme functional. It is therefore necessary to reformulate the reduced-matrix-element-dependent form-factors (\ref{ff_sph}) to a body-fixed frame. Usual derivation \cite{FW66} starts with a fixed $\vec{q}$-direction, which translates to fixed $\mu$ (usually through the outer sum $\sum_\mu\mathscr{D}_{1\mu}^\lambda(\hat{q})\ldots$).
\begin{align}
\frac{1}{2J_i+1}\sum_{M_i M_f}\frac{\big|\langle J_f||\hat{T}_\lambda||J_i\rangle\big|^2}{2J_f+1}
\Big(C_{J_i M_i \lambda\mu}^{J_f M_f}\Big)^2
&= \frac{\big|\langle J_f||\hat{T}_\lambda||J_i\rangle\big|^2}{(2J_i+1)(2\lambda+1)} \\
=\frac{\big|\langle J_f||\hat{T}_\lambda||J_i\rangle\big|^2}{(2J_f+1)(2\lambda+1)}
\sum_{\mu,M_f}\Big(C_{J_i M_i \lambda\mu}^{J_f M_f}\Big)^2
&= \sum_{\mu,M_f} \frac{\big|\langle J_f M_f|\hat{T}_{\lambda\mu}|J_i M_i\rangle\big|^2}{2\lambda+1}
\end{align}
The bottom line shows that the same result can be achieved from the body-fixed point of view through fixed $M_i$ and summed $\mu,M_f$; and this sum, in fact, doesn't depend on $J_i,\,J_f$ (which are undefined in axial symmetry). The equivalent replacement in the spherical form-factors is therefore $\frac{1}{2J_i+1}\big|\langle f||\hat{T}_\lambda||i\rangle\big|^2 \ \mapsto\ \sum_{\mu,M_f} \big|\langle f|\hat{T}_{\lambda\mu}|i\rangle\big|^2$, and for even-even nuclei we use
\begin{equation}
F_{\lambda;fi}^{\mathrm{(ax)}} = \langle f |\hat{T}_{\lambda K}|i\rangle \times\bigg\{
\begin{array}{cl} \!\!1 & \textrm{for }K=0 \\ \!\!\sqrt{2} & \textrm{for }K>0. \end{array}
\end{equation}

\subsection{Other known corrections}\label{sec:corr}
Finally, let us say a few words about the already well-known refinements. The recoil term $f_\mathrm{rec}$ in (\ref{crosssec}) comes from the integration of energy-conserving delta-function $\delta(\sum E)$ with the radial phase space of the scattered electron. Moreover, as omitted in \cite{FW66}, but included in \cite{GrQED}, we need to integrate also the momentum-conserving delta-function $\delta^3(\sum \vec{p})$ in the phase space of recoiled nucleus, which only adds a close-to-one factor $\tfrac{Mc^2+\hbar\omega_\nu}{E_n}$.
\begin{equation}
f_{\mathrm{rec}} = \frac{Mc^2+\hbar\omega_\nu}{E_n + 2E_f\sin^2\tfrac{\theta}{2} - \tfrac{E_f\Delta p}{p_f}\cos\theta}
\approx \frac{Mc^2 + \hbar\omega_\nu}{Mc^2+2E_i\sin^2\tfrac{\theta}{2}}
\end{equation}

\begin{figure}[htb]
\centerline{%
\includegraphics[width=12.5cm]{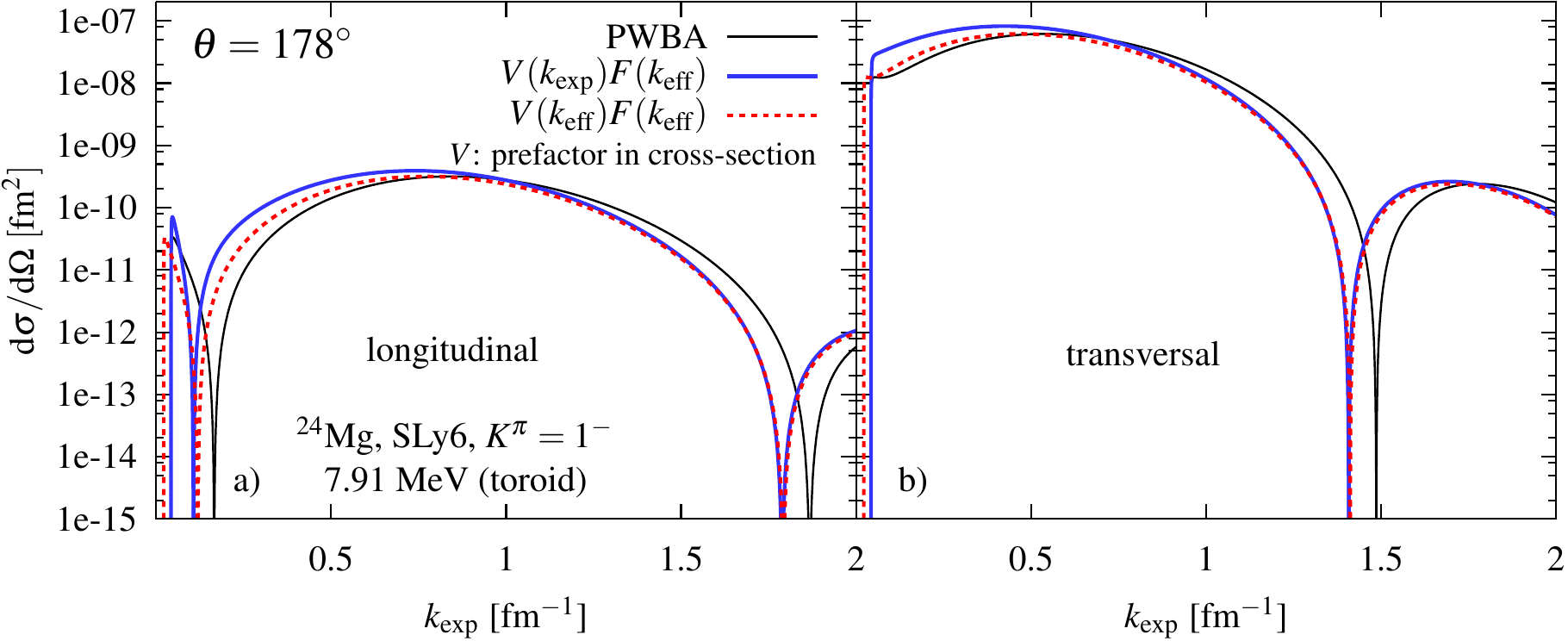}}
\caption{The impact of effective-momentum correction, which should mimic the more sophisticated DWBA approach, which we didn't implement yet. It is not clear whether $k_{\mathrm{eff}}$ should also be used in prefactors, so we are showing both options.}
\label{Fig:keff}
\end{figure}

Finally, all the above-mentioned methods relied on the plane-wave Born approximation (PWBA). In order to mimic the more appropriate DWBA, we should evaluate form-factors at the effective momentum \cite{FW66,HB83}
\begin{equation}
k_{\mathrm{eff}} = k_{\mathrm{exp}}\bigg(1+\frac{3}{2}\frac{Z\alpha\hbar}{p_i R}\bigg)\qquad
\textrm{with }\ R=1.12A^{1/3}\ \mathrm{fm}.
\end{equation}
However, it is not clear which momentum should be employed in the evaluation of prefactors in (\ref{crosssec}). Both cases are demonstrated in Fig.~\ref{Fig:keff} for backward scattering, which shows relatively consistent shift of the nodal points. In forward scattering, the differences among all cases are proportionally smaller, and are not shown here.

\section{Conclusions}
Momentum dependence of the inelastic-electron-scattering cross section can provide valuable information about the shape of transition densities and currents. The main (but non-specific) effect of toroidal/compression flow is the decreasing amplitude at low $k$, and a further case-study is still necessary for a firm assignment. Transversal form-factors, which reflect the transition current, become dominant for the back-scattering geometry (e.g., $\theta=178^\circ$).

The influence of various corrections was investigated. Effective current affects transversal cross-section of isovector transitions mainly by overall scaling of around 25\%. Exact relativistic kinematics affects the low-$k$ part of back-scattering longitudinal cross section, which may be important for correct error-analysis, in the case when longitudinal/transversal separation is not done independently, e.g.~by angular correlations from $(e,e'\gamma)$. Effective-momentum approximation (or the full DWBA) shifts the nodal points in the momentum space, which affect the spatial details of the reconstructed transition density/current flow. Besides, a proper numerical calculation of form-factors should not omit the elimination of the spurious admixtures.
\vskip2ex
This work was supported by Slovak Research and Development Agency (Contract No.~APVV-15-0225) and Votruba-Blokhintsev (Czech Republic--BLTP JINR) grant. A.R.~thanks Valentin O. Nesterenko from BLTP JINR Dubna for fruitful discussions and for providing relevant references.

\end{document}